\documentclass[prl,showpacs,twocolumn,typeset,superscriptaddress]{revtex4} 

\usepackage{times,txfonts} 
\usepackage{dcolumn}
\usepackage{bm}
\usepackage{verbatim}
	
\begin{document}
\title{Reply to ``Comment on `Witnessed entanglement and the geometric measure of
 quantum discord'  "}  
\author{Tiago Debarba} 
\email{debarba@fisica.ufmg.br}
\author{Thiago O. Maciel}
\author{Reinaldo O. Vianna}

\affiliation{Departamento de F\'{\i}sica - ICEx - Universidade Federal de Minas Gerais,
Av. Pres.  Ant\^onio Carlos 6627 - Belo Horizonte - MG - Brazil - 31270-901.}

\date{\today}
\begin{abstract}

We show that the mistakes pointed out by    Rana and Parashar 
[Phys. Rev. A {\bf 87}, 016301 (2013)] do not invalidate the main conclusion
of our work [Phys. Rev. A {\bf 86}, 024302 (2012)]. 
We show that the errors  affected only a particular 
application of our general   results, and present the correction.

\end{abstract}
\maketitle

 Rana and  Parashar \cite{comment}   claim that our 
bounds between geometrical discord and entanglement
\cite{debarbapra} are incorrect. They give examples of violations of our
bounds and suggest it has to do  with  non-monotonicity 
of geometrical discord in the Hilbert-Schmidt norm. 
The authors started their comment revising our definition of geometrical discord and 
pointing a typographical error in the definition of negativity. We defined negativity
as  the sum of the negative eigenvalues of the partial transpose of 
the state,  Eq.16 of our work, while some authors further normalize this quantity.
Their critique about  the normalization of 
the geometrical discord in the Hilbert-Schmidt norm is also irrelevant, 
for the normalized geometrical discord is greater than ours.

The first counterexample which would violate our results is the 
maximally entangled state for two qubits ($\phi_{+}$). They 
consider the negativity
as $1$, while the  2-norm geometrical 
discord is $1/2$. 
But it is not correct. Consider Eq.20 ,
\begin{equation}
D_{(2)}(\phi_{+})\geq \frac{E_{w}^2}{Tr(W_{\phi_{+}}^2)}.
\end{equation}
We have $D_{(2)}(\phi_{+})=1/2$, and 
$E_{w}=Tr(W_{\phi_{+}}\phi_{+})=Tr(P_{-}\phi_{+}^{T_1})=1/2$, where $P_{-}$ is the 
projector associated to the negative eigenvalue  of the partial transpose of $\phi_{+}$. 
$Tr(W_{\phi_{+}}^2)$ is the number of negative eigenvalues of the partial transpose, which
is $1$. 
Thus  $D_{(2)} = 1/2 \geq E_{w}^2=1/4$.

The next counterexample is the $2\otimes 32$-dimension state. For this example we have 
quantum discord $D_{(2)}(\rho)=0.01$ and $E_{w}^2/Tr[W_{\rho}^2]=0.0032$, where $E_{w}$ 
is the negativity, and  Eq.20 is satisfied. However, in the 
comment the equation taken was Eq.21, and via that relation we get
 $\mathcal{N}^2/(d-1)^2=0.0316$,
which violates the bound. The point is we mistakenly had written that 
$Tr[W_{\rho}^2]\leq d-1$, for a system with dimension $d\otimes d'$ and $d\leq d'$. 
In the counterexample  we have $Tr[W_{\rho}^2]=10$, i.e. the partial transpose of the 
state has $10$ negative eigenvalues  and 
not $d-1=1$, and this  is the reason of the wrong violation in Eq.21. 
In the comment, the authors  conclude that the violation comes from the fact 
that $D_{(2)}(\rho)$ is not a monotonic distance, but monotonicity does not play 
any role in our bounds.

Finally,  the authors claim that Eq.27 is not valid. 
Equation 27 is a particular case of 
Eq.22, where we get a linear relation  between geometrical discord  calculated via trace norm 
and witnessed entanglement. This bound is valid only  for entanglement measures whose optimal 
entanglement witnesses live in the domain $-\mathbb{I} \leq W \leq \mathbb{I}$,
and the entanglement witness for the negativity is not in this domain, which explains the 
problem with
the bound in Eq.27. An example of entanglement measure for which 
this bound is valid is the random robustness of entanglement, Eq.28. 
Equation 27 can be easily corrected by means of an inequality more general than Eq.22, namely:
\begin{equation} 
D_{(1)}\geq \frac{E_w}{|| W_{\rho}||_{\infty}}, 
\end{equation}
where $|| W_{\rho}||_{\infty}$ is the greatest eigenvalue of the optimal entanglement witness 
of the state $\rho$
\footnote{Take the well known inequality 
for  operators $A$ and $B$,  
$||A||_{q}||B||_{p}\geq |Tr[AB^{\dagger}]|$, for $1/q+1/p = 1$. Set
$A=\rho-{\xi}$, where $\xi$ is $\rho$'s nearest non-discordant state, and set $B = W_{\rho}$, 
where $W_{\rho}$ is the optimal entanglement witness of $\rho$, then follows straightforwardly
$D_{(p)} \geq E_{w}/||W_{\rho}||_{q} $. For $p=1$, we have $q=\infty$.}. 
 Note that  this bound is valid for every witnessed entanglement.

In conclusion, the main results of our work are  Eq.20 and Eq.22,  which are rigorously correct.
They were  calculated from 
{\it first principles},
via well known  inequalities for operators and properties of entanglement witnesses.
We made two mistakes when specializing  for the 
negativity, as discussed and clarified above. 
The conjecture proposed by D. Girolami 
and G. Adesso \cite{girolamiadesso} about the interplay between geometrical 
 quantum discord and  entanglement is implicit
in  Eq.20 and Eq.22.

\bibliographystyle{apsrev}

\end{document}